\documentstyle[amssymb,preprint,eqsecnum,aps,prb,amstex]{revtex}

\begin{document}
\title{Ca$_{0.85}$Sm$_{0.15}$MnO$_3$: A mixed antiferromagnet with unusual
properties}
\author{R. Mahendiran, A. Maignan, C. Martin, M. Hervieu, and B. Raveau}
\address{Laboratoire CRIMAT, ISMRA, Universit\'{e} de Caen, 6. Boulevard du\\
Mar\'{e}chal Juin, Caen Cedex -14050, France}
\date{\today}
\maketitle

\begin{abstract}
We establish a phase diagram for the electron-doped manganites Ca$_{1-x}$Sm$%
_x$MnO$_3$ ( 0 $\leq $ x $\leq $ 0.15). The low temperature insulating phase
of x = 0.15 is a mixed antiferromagnet with two long range
antiferromagnetism, C-type (monoclinic) and G-type (orthorhombic),
coexisting with short range ferromagnetic clusters (orthorhombic).
Resistivity ($\rho $) and magnetization ($M$) of x = 0.15 show unusual
magnetic field history dependent phenomena which are not observed for x $%
\leq $ 0.12: irreversibilty between zero field- cooled (ZFC) and
field-cooled (FC) data below 120 K and hysteresis between cooling and
warming for all values of magnetic fields (0 $\leq $ H $\leq $ 7 T). Field
cooling strongly enhances M (M$_{FC}$/M$_{ZFC}$ = 2.7 at H = 5 T and 10K)
reduces $\rho $ ($\rho _{FC}$/$\rho _{ZFC}$ =1.5$\times $10$^{-4}$ at 7 T
and 10 K) and even induces metallic-like resistivity (d$\rho $/dT 
\mbox{$>$}%
0) for H = 7 T below 80\ K. We discuss the possible origins of the results.
\end{abstract}


\smallskip

Coexistence of itinerant and localized charges over different length scales,
known as the electronic phase separation, seems to be one of the fundamental
aspects of colossal magnetoresistive manganites of the type R$_{1-x}$A$_x$MnO%
$_3$ where R and A are trivalent rare earth and divalent alkali ions
respectively. The electronic phase separation in manganites also induces
magnetic phase separation as the hopping of e$_g$ hole between Mn$^{4+}$:t$%
_{2g}^3e_g^0$-O-Mn$^{3+}$:t$_{2g}^3$e$_g^1$ is facilitated if t$_{2g}^3$
spins are ferromagnetically aligned and hindered if they are
antiferromagnetically aligned. Thus, phase separation manisfests itself as
isolated ferromagnetic polarons or clusters with itinerant charges in either
antiferromagnetic insulating matrix or paramagnetic insulating matrix or
random mixtures of ferromagnetic metallic and antiferromagnetic insulating
domains of various sizes. Some of the experimental evidences are: mobile
ferromagnetic droplets in the antiferromagnetic La$_{1-x}$Ca$_x$MnO$_3$ (x =
0.08, 0.1)\cite{Hennion}, few ten angstrom size ferromagnetic clusters in
the paramagnetic insulating phase of La$_{0.67}$Ca$_{0.33}$MnO$_3$ -type
compounds \cite{Teresa} nanometer to micron size ferromagnetic clusters
within the charge ordered matrix of La$_{0.5\pm \delta }$Ca$_{0.5\pm \delta
} $MnO$_3$, Nd$_{0.5\pm \delta }$Sr$_{0.5\pm \delta }$MnO$_3$ and (PrLa)$%
_{0.7} $Ca$_{0.3}$MnO$_3$\cite{Mori}. Although some recent theoretical models%
\cite{Nageav} predict phase separation of a few angstrom size, micron size
domains found experimentally appears to be connected with structural phase
separation\cite{Ritter}.

Most of the existing reports are on hole-doped (Mn$^{3+}$rich or x $\leq 0.5$%
) compounds \cite{Hennion,Teresa,Mori}$.$ An interesting type of phase
separation occurs in the electon-doped Ca$_{0.85}$Sm$_{0.15}$MnO$_3$. It is
paramagnetic and single phase with orthorhombic (Pnma) structure at 300 K,
but it phase separates into {\it two} {\it long range} {\it antiferromagnetic%
} phases, G- and C- types below 130 K and coexist with each other in
orthorhombic ({\it Pnma}) and monoclinic (P2$_1$/m) structures respectively
down to 5 K\cite{Martin1}. Interestingly, this particular composition of
mixed antiferromagnet showed the highest magnetoresistance in the series Ca$%
_{1-x}$Sm$_x$MnO$_3$\cite{Martin2}$.$ The origin of colossal
magnetoresistance in this compound is not understood yet. In this
communication we bring out anomalous magnetic field history dependent
behavior of resistivity and magnetization in x = 0.15 and contrast our
results with lower doping ($x$) levels. We also establish the magnetic phase
diagram for the first time.

\smallskip

Four probe resistivity measurements in the temperature range from 300 K to 5
K up to the maximum field of H = 7 T was done using Quantum Design Physical
Property Measuring system. Magnetization up to H = 5 T was done using
Quantum Design SQUID magnetometer in the temperature range 300 K-5 K.
Measurements were done in three methods: in the zero field cooled (ZFC) and
field cooled (FC) modes, the sample was cooled from 300 K to 5 K rapidly (10
K/min) in absence of external magnetic fields and in presence of a known
field (H) respectively and the data were taken while warming (2 K/min) from
5 K. In the thermal cycling under magnetic field (TCUF) mode, the sample was
subjected to a known field (H) at 300 K and the data were collected while
cooling down to 5 K and warming back to 300 K at a rate of 2 K/min.

\smallskip

\smallskip Fig. 1(a) shows the phase diagram of Ca$_{1-x}$Sm$_x$MnO$_3$
obtained from the low temperature magnetization (Fig. 1(b)) and resistivity
data. The spontaneous magnetization, M(0T), obtained from the linear
extrapolation of high field data to H = 0 T and the high field
magnetization, M(5T) from Fig. 1(b) show similar trend with x: a rapid
increase in between x = 0.05 and x = 0.075, a maximum around x = 0.12 and a
reduced value at x = 0.15. Even though M(H) of x = 0.075-0.12 at low fields
resembles a long range ferromagnet, M increases continuoulsy without
saturation at higher fields. The magnetization at H = 5 T, M(5T), is far
below the value M(F) for the fully aligned t$_{2g}^{_{^3}}$ and e$_g^1$
spins. This important observation lead us to suggest a heterogeneous
magnetic state in Ca$_{1-x}$Sm$_x$MnO$_3$ as illustrated by the schematic
diagrams in Fig. 1(a). Region I (0 $<$ x $\leq $ 0.05) is characterized by
ferromagnetic (FM) clusters (black circles) embedded in a uniform G-type
antiferromagnetic (G-AF) background (hatched region). These ferromagnetic
clusters are created by the polarization of Mn$^{4+}$ (t$_{2g}^3$) spins
around the doped Mn$^{3+}$ (t$_{2g}^3$e$_g^1$) ions by double exchange
interaction and doped charges (e$_g$ electrons) are itinerant within these
clusters. The onset of ferromagnetic order within these clusters sets in at T%
$_C$ = 118$\pm 3$ K as determined from low field susceptibility measurements
and scarcely varies with doping level x. As x increases, FM clusters size
increase and they percolate in region II (0.05 $<$ x $<$ 0.13) still in G-AF
background. Region III (0.13 $\leq $ x$\leq $ 0.15) is dominated by C-type
magnetic order but smaller G-AF and FM clusters coexist. These changes in
magnetic properties are also reflected in electrical resistivity at 5 K ($%
\rho ($5K$)$) which decreases by 5 orders of magnitude from x = 0 to x =
0.12 and increases again as the antiferromagnetic order changes to C-type.
The samples in regions II show metallic like resistivity (d$\rho $/dT 
\mbox{$>$}%
0) below 100 K due to the percolation of FM clusters. The composition of our
primary interest is x = 0.15 which shows C-AF ordering in monoclinic
structure below T$_{NC}$ $=$112 K and coexist with orthorhombic FM (T$_C$ $%
\thickapprox $ 118 K) G-phases (T$_{NG}$ $\thickapprox $ 118 K).

\smallskip

Fig. 2(a) shows the resistivty $\rho $(T) recorded under the TCUF mode. As T
decreases from 300 K, $\rho $(0T) initially decreases linearly with T down
to 200 K, shows a minimum around T$_p$ = 160 K and increases again below
this temperature as shown by the enlarged view in the inset. However, a
rapid increase in $\rho $(T) occurs at still lower temperature, T$_{NC}$ =
112 K and changes by 4 orders of magnitude as T lowers to 5 K. The data
taken during warming from 5 K bifurcate from the cooling curve and maintains
higher resistivity values in the temperature range 70 K-125 K suggesting the
first order nature of the transition. The rapid decrease in $\rho $(T)
around 125 K while warming closely correlates with disappearance of the
C-type AF magnetic order as found by neutron diffraction\cite{Martin1}. We
find a large reduction in $\rho $(T) below 115 K for various values of H,
but a metallic- like resistivity behavior (d$\rho $/dT $>$ 0) is seen only
at H = 7 T below 80 K. We measured $\rho $(T) at H = 6 T also (not shown
here for clarity) but d$\rho $/dT was found to be negative below 110 K. $%
\rho $(T) under different values of H show hysteresis of nearly same width
as in H = 0 T data. The increase in $\rho $(7T) just below 102 K is possibly
related to the shift of T$_{NC}$ from 112 K for H = 0 T to 102 K for H = 7
T. On the high temperature side, the resistivity minimum at T$_p$ is
gradually suppressed with increasing H as shown in the inset.

\smallskip

The unexpected magnetic field history dependence of $\rho $(T) is shown in
Fig. 2(b). The FC resistivity curves (dashed lines) recorderd while warming
from 5 K are similar to those ones in Fig. 2(a). However, the ZFC curves
(thick lines) are distinctively different: they are higher in resistivities
than their FC counterparts and the curves under different H closely resemble
the temperature dependence of $\rho $(0T) itself. It should be noted that
while the field cooled $\rho $(7T) decrease continuously with temperature
below 70 K, the decrease of zero field cooled $\rho $(7T) below 75 K is
overwhelmed by a resistivity upturn below 35 K. The resistivity ratio $\rho
_{FC}$/$\rho _{ZFC}$ between zero field cooling and field cooling is as
small as 1.5$\times $10$^{-4}$ at 10 K and 7 T. These differences are found
only in samples close to x = 0.15 (0.13 $\leq $ x $\leq $ 0.15 but {\it not}
in any other compositions for x $\leq $ 0.12 as shown in the inset of Fig.
2(b) for the insulating compound x = 0.025.

\smallskip

Motivated by the above unusual magnetotransport results and keeping in mind
that magnetotransport in these materials are sensitive to the underlying
magnetic order, we investigated the field and temperature dependence of the
magnetization (M) in details as shown in Fig. 3(a) and 3(b) corresponding to
Fig. 2(a) and 2(b) respectively. The maximum of M under 0.01 T at T$_{NC}$ =
112 K (see Fig. 3(a)) while field cooling signals the onset of simultaneous
C-type antiferromagnetism and orthorhombic({\it Pnma}) to monoclinic(P2$_1$%
/m) transformation. The phase fraction of monoclinic phase increases from 64 
$\%$ at 110 K to 94 \% at 10 K\cite{Martin1}. We find that M(T) curve while
field heating deviates from the field cooling branch starting from 70 K for
H = 5 T (90 K for H = 0.01 T), keeps a value lower than the field cooled
ones, reaches a maximum at about 2 K above than while cooling and merges
with the field cooling curve above 120 K. This trend in M(T) is also
reflected in $\rho $(T) in Fig. 2(a) which shows higher value of $\rho $
while warming than cooling. These hysteresis behaviors in $\rho $(T) and
M(T) are the consequence of first order magneto-structural transition
involving nucleation of high resistance, C-type antiferromagnetic monoclinic
phase in low resistance, paramagnetic orthorhombic matrix while cooling and
vice versa on heating from low temperature. In concurrence with the
resistivity behavior in Fig. 2(b), a large difference between ZFC (symbols)
and FC (thick lines) magnetization occurs below 120 K (see Fig. 3(b)) and
the diffference increases with increasing H and decreasing T. No difference
between FC and ZFC magnetizations for H $\geq $ 1 T is found for x $\leq $
0.12.

\smallskip \smallskip

Fig. 4 allows us to have further insight into the magnetic behavior observed
above. Field cycling (0 T$\rightarrow $1T$\rightarrow $ -1T$\rightarrow $1T)
data recorded at 10 K after zero field cooling (marked 1 T(ZFC)) shows a
rapid increase in M for less than 20 Oe and reaches a maximum value of 0.136 
$\mu _B.$ However, M increases by 25 \% at 1 T when the sample is field
cooled (marked as 1 T(FC)) from T 
\mbox{$>$}%
T$_{NC}$ (125 K). A large enhancement in M under field cooling is clearly
seen for all the measured values of H and we do not find hysteresis in M up
to 2 T. The right inset of Fig. 4 compares M cooled under 5 T to the zero
field-cooled curve at 5 K. For H = 4 T (main panel) and 5 T (right inset), M
has higher values while decreasing H from its maximum value H$_{max}$ to 0
T, but on subsequent field cyling (0 T$\rightarrow -$H$_{max}\rightarrow +$H$%
_{max}$) M settles to slightly lower values. This behaviour is {\it not}
caused by time dependent decay of magnetization since the data were recorded
5-10 minutes after the stablization of temperature. M at H = 5 T (in the
virgin field-cooled curve) is enhanced by a factor of with respect to zero
field-cooled value at 5 T (M$_{FC}$/M$_{ZFC}$ = 2.7). The observed
enhancement of magnetization occurs {\it only} if the sample is field cooled
from T 
\mbox{$>$}%
T$_N$ and not if T 
\mbox{$<$}%
T$_N$. We find similar trends in 0.13 $\leq $ x $\leq $ 0.15 but do not
observe in other compositions (x 
\mbox{$<$}%
0.12) as shown for x = 0.025 in the left inset.

\smallskip

The surprising magnetic field history dependent properties found for x =
0.15 but not for x = 0.025 (or x $\leq $ 0.12) are difficult to understand
from view point of magnetic heterogeneity alone because it prevails in both
(and in all) these compounds. The increasing value of low field magnetic
moments under field cooling for increasing strength of H suggests that more
and more spins are getting aligned with H and the ferromagnetic clusters
increase in size. It is unlikely that spins in the G-type AF phase
contribute to this behavior because such trends lack for x $\leq $ 0.12.
Since the enhancement of M is found {\it only} when the sample is field
cooled from T 
\mbox{$>$}%
T$_N$ ( = T$_{S,}$ the transition temperature for structural transition),
spins in the interfacial region between monoclinic C-type AF and
orthorhombic FM phase might play important role. A pronounced increase in
field cooled magnetization even at high values of H was first discovered for
ferromagnetic nanoparticles of Co covered with antiferromagnetic CoO layers 
\cite{Bean} and studied extensively in recent times in connection with
exchnage anisotropy/exchange biasing between ferromagnetic-antiferromagnetic
films\cite{Berko,Maloz}. Field cooling Co/CoO mixures from T 
\mbox{$>$}%
T$_C$ ( where T$_C$ is the ferromagnetic Curie temperature of Co) aligns
magnetic moments of single domain Co particles in the field direction but
certain fraction of spins of antiferromagnetic CoO at the interface are
exchange coupled to Co moments which aligns themselves with Co spins.
However, hystersis loop of such exchange coupled systems made under FC mode
are shifted from the origin which we do not see in our compounds. It does
not mean that exchange coupling is not playing role in our compound.
Existing theoretical model\cite{Maloz,Berko} assume that there are no
macroscopic structural changes under external magnetic fields on either side
of the interface. But, there are clear evidences that field induced
structural changes accompany antiferro to ferromagnetic transition in
manganites as we have shown for Nd$_{0.5}$Sr$_{0.5}$MnO$_3$.\cite
{Mahi1,Ritter} This compound is also structurally and magnetically phase
separated with minority long range ferromagnetic orthorhombic phase
coexisting with majority charge ordered antiferromagnetic monoclinic phase
at low temperatures and we also find magnetic history dependent behavior
similar to Ca$_{0.85}$Sm$_{0.15}$MnO$_3$ as shown in Fig. 5. Our neutron
diffraction study under magnetic fields in Ca$_{0.85}$Sm$_{0.15}$MnO$_3,$
although not done in a systematic way as done here, confirms monoclinic to
orthorhombic tranformation whose fraction also depends on the temperature
and the strength of external magnetic field\cite{Mahi2}.

\smallskip \smallskip

In the absence of any theoretical models dealing with such exchange biasing
relevant to manganites, we borrow our ideas from the random field model of
Imry and Ma \cite{Imry} which was applied to variety of different systems
including exchange anisotropy/bias systems\cite{Maloz}, Ising
antiferromagnets with random impurities\cite{Grest,Belanger}, mixed Ising
Jahn-Teller system\cite{Graham} DyV$_{1-x}$As$_x$O$_4$ and martensitic
transformation\cite{Katch} and relaxor ferroelectrics\cite{Kleemann}. The
basic idea of Imry and Ma\cite{Imry} is that systems in which random field
effects dominates, domain formation is energetically favoured over long
range order. Experimental realization of random field effect in an Ising
antiferromagnet with random impurities is obtained under field- cooled
condition\cite{Grest,Belanger}. Ca$_{0.85}$Sm$_{0.15}$MnO$_3$ with C-type
antiferromagnet in which successive ferromagnetic linear chains along c-axis
are coupled antiferromagnetically is an Ising antiferromagnet. The C-type
antiferromagnetic phase also exhibit cooperative Janh-Teller distortion due
to ordering e$_g$-d$_Z2$ orbitals along c- axis\cite{Martin1}. The quenched
random impurities are the G-type and FM phases. In zero field cooled
measurement, the long range order corresponds to the coexistence of majority
C-type AF phase and minority G-type AF phase and FM phases. The resistivity
of zero field -cooled state is high as electron hopping between
antiferromagnetically coupled Mn$^{3+}$ and Mn$^{4+}$ sites is not favoured
by double exchange interaction. Field cooling enhances the size of
ferromagnetic regions and, breaks the C-type AF matrix into domains due to
random field effect\cite{Imry,Grest,Belanger}. Then, domain walls in which
spins are not exactly antiparallel along with the expanded FM regions in the
orthorhombic phase constitute least resistance path for electrical
conduction and resistivity decreases. As the external field increases above
4 T, a partial structural transformation from monoclinic, C-type
antiferromagnetic to orthorhombic (ferromagnetic) also takes place. Our
neutron diffraction results\cite{Mahi2} suggest that a complete monoclinic
to orthorhombic structural transformation occurs at T = 100 K and H = 6 T
but the transformation is only partial at low temperatures(60 \% monoclinic
and 40 \% orthorhombic at 40 K and H = 6 T after a zero field cooled
process). Hence, when the sample is field cooled for H $\geq $ 4 T, the
magnetic moment is initially high (see the inset of Fig. 4) since the
orthorhombic ferromagnetic phase contributes largely to the observed
magnetization but its fraction decreases as H is reduced to zero. Upon,
further field cyling from 0 T $\rightarrow -5T\ \rightarrow 5T$
magnetization locks to a value determined by the new phase fraction of
orthorhombic/monoclinic phases. Thus, the field cooling is more efficient in
reducing the resistivity than the zero field cooling. The absence of Ising
spin character and structural variants in lower doped compounds is the most
likely the reasons why we fail to observe magnetic field history dependent
behaviors for lower x.

\smallskip

In conclusion, resistivity and magnetization of the electron doped compound
Ca$_{0.85}$Sm$_{0.15}$MnO$_3$ reveals first order nature of
paramagnetic-antiferomagnetic transition and field cooling enhances
magnetization, reduces resistivity and even induces insulator-metal
transition for H = 7 T whereas it is an antiferromagnetic insulator when
cooled in zero field. No difference between field cooled and zero field
cooled resistivities are found for x $\leq $ 0.12. These differences are
suggested to the mixed phase (two antiferromagnetic phases and a
ferromagnetic phase coexisting in two different crystallographic structures)
nature of Ca$_{0.85}$Sm$_{0.15}$MnO$_3$ in zero field and formation of AF
domain states and structural transition under magnetic fields.

\smallskip

R.M thanks MNERT (France) for financial support and acknowledges discussions
with Professors T. V. Ramakrishnan, M. R. Ibarra, D. I. Khomskii and Drs.
Venkatesh Pai, C. Ritter and P. A. Alagarebel.

\newpage FIGURE CAPTIONS

\begin{description}
\item[Fig.1]  : (a). Phase diagram of Ca$_{1-x}$Sm$_x$MnO$_3$ ( 0 $\leq $ x $%
\leq $ 0.15). $\rho ($5K$):$ Resistivity at 5 K, $M$(5 T): Magnetization at
H = 5 T and at 5 K. M (0T): Extrapolation of high field M to H = 0 T. M(F):
Expected magnetic moments for fully ferromagnetically aligned t$_{2g}$, e$_g$
spins. Lines are guide to the eyes. Black circle: Ferromagnetic clusters,
Hatched regions: G- and C- type antiferromagnetic phases. (b). Field
dependence of magnetization at 5 K for Ca$_{1-x}$Sm$_x$MnO$_3$.

\item[Fig.2]  :(a). Resistivity ($\rho $) of x = 0.15 recorded during
thermal cylcing under magnetic field (TCUF). Arows indicate the direction
temperature sweep. Inset: expanded view of $\rho $ above 100 K. (b). $\rho
(T)$ made under zero field-cooled (thick lines) and field-cooled (dashed
lines). Inset: $\rho (T)$ of x = 0.025. Double head arrows are to indicate
perfect reversibility. Note that there is no difference in $\rho $(T)
between zero field cooling and field cooling conditions.

\item[Fig.3]  : (a). Magnetization (M) of x = 0.15 made under the TCUF mode.
(b). M under ZFC (symbols connected by lines), FC (thick lines) modes.

\item[Fig.4]  : Main panel: Magnetic field cylcing of magnetization made
under field cooled (FC) mode ( +H$_{max}$ $\rightarrow $ -H$%
_{max}\rightarrow +$H$_{max}$) for Hmax = 1 T, 2T, 3 T, 4T. Magnetization
under zero field cooled (ZFC) mode (0 T $\rightarrow $ +Hmax$\rightarrow $
-Hmax$\rightarrow $ +Hmax) is also shown for 1 T. Double head arrows are to
indicate the complete reversibility. Right inset : M under ZFC and FC modes
up to H = 5 T. Left inset: M of x = 0.025. Note that there is no difference
in M between FC and ZFC mode.

\item[Fig.5]  : Resistivity of Nd$_{0.5}$Sr$_{0.5}$MnO$_3$ made under zero
field-cooled (thick lines) and field-cooled (dashed lines) modes. Arrows
indicate the direction of temperature sweep.
\end{description}


\begin{references}
\bibitem{Hennion}  M. Hennion, F. Moussa, J. Rodriguez-Carvajal, L. Pinsard,
and A. Revcolevschi, Phys. Rev. {\bf B 56}, R497 (1997).

\bibitem{Teresa}  J. M. de Teresa, M. R. Ibarra, C. Marquina, P. A.
Algarebel, C. Ritter, J. Garcia, J. Blasco, A. del Morel, and Z. Arnold,
Nature {\bf 386}, 256 (1997); J. Z. Sun, L. Krusin-Elbaum, A. Gupta, and G.
Xiao, Appl. Phys. Lett. {\bf 69}, 1002 (1996).

\bibitem{Mori}  S. Mori, C. H. Chen, and S. W. Cheong , Phys. Rev. Lett. 
{\bf 81} 3972 (1998); R. Mahendiran, M. R. Ibarra, A. Maignan, A. Arulraj,
R. Mahesh, B. Raveau, and C. N. R. Rao, Phys. Rev. Lett. {\bf 82}, 2191
(1999); R. Mahendiran, M. R. Ibarra, A. Maignan, C. Martin, B. Raveau, and
A. Hernando, Solid State Commun. {\bf 111}, 525 (1999); G. Allodi, F. Licci,
and W. Pepper, Phys. Rev. Lett. {\bf 81} 4736 (1999); M. Roy, P. Schiffer,
and A. P. Ramirez, Phys. Rev. B {\bf 58}, 5158 (1999); M. R. Ibarra, G. M.
Zhao, J. M. de Teresa, Z. Arnold, C. Marquina, P. A. Algerabel, H. Keller,
and C. Ritter, Phys. Rev. B {\bf 57}, 7446 (1998); N. Babuskina, L. M.
Belova, D. I Khomskii, K. I Kugel, O. Yu. Gorbekko, and A. R. Kaul, Phys.
Rev. B {\bf 59}, 6994 (1996).

\bibitem{Nageav}  E. L. Nagaev, Physics Uspekhi {\bf 39}, 781 (1996); A.
Moreo, S. Yunoki, E. Dagoto, Science, {\bf 283}, 2034 (1999); M. Yu. Kagan
and D. I. Khomskii, Euro. Phys. J. B {\bf 12}, 217-223 (1999); L.P. Gork'ov
and V. Z. Kresin, JETP Lett. {\bf 67}, 985 (1998)

\bibitem{Ritter}  C. Ritter, R. Mahendiran, M. R. Ibarra, A. Maignan, B.
Raveau, and C. N. R. Rao, Phys. Rev. B 61, R9229 (2000).

\bibitem{Martin1}  C. Martin, A. Maignan, M. Hervieu, B. Raveau, Z. Jirak A.
Kurbakov, V. Tronouv, G. Andre, and F. Bouree, J. Magn. Magn. Mater. {\bf 205%
}, 184 (1999).

\bibitem{Martin2}  C. Martin, A. Maignan, F. Damay, M. Hervieu, and B.
Raveau, J. Solid State Chem. {\bf 134}, 198 (1997).

\bibitem{Bean}  W. H. Meiklejohn and C. P. Bean, Phys. Rev. {\bf 102}, 1413
(1956).

\bibitem{Berko}  For review see, A. E. Berkowitz and K. Takano, J. Magn.
Magn. Mater. {\bf 200}, 552 (1999)

\bibitem{Maloz}  A. P. Malozemoff, J. Appl. Phys. {\bf 63}, 3874 (1988).

\bibitem{Mahi1}  R. Mahendiran, M. R. Ibarra, A. Maignan, F. Millange, A.
Arulraj, R. Mahesh , B. Raveau, and C. N. R Rao, Phys. Rev. Lett. {\bf 82},
2191 (1999).

\bibitem{Mahi2}  R. Mahendiran, M. R. Ibarra, P. A. Algarebel, A. Maignan,
C. Martin, B. Raveau, and C. Ritter (unpublished).

\bibitem{Imry}  Y. Imry and S. Ma, Phys. Rev. Lett. {\bf 35}, 1399 (1975).

\bibitem{Grest}  G. S. Grest, C. M. Soukoulis, and K. Levin, Phys. Rev. {\bf %
B} {\bf 33}, 7659 (1986).

\bibitem{Belanger}  For review see, D. P. Belanger in {\it Spin glass and
Random fields}, World Scientific (Singapore 1997).

\bibitem{Graham}  J. T. Graham, M. Maliepaard, J. H. Page, S. R. P. Smith,
and D. R. Taylor, Phys. Rev. {\bf B 35}, R2098 (1987).

\bibitem{Katch}  S. Semenovskaya and A. G. Khachturyan, J. Appl. Phys. {\bf %
83}, 5125 (1998).

\bibitem{Kleemann}  V. Westphal, W. Kleemann, and M. D. Glinchuk, Phys. Rev.
Lett. {\bf 68}, 847 (1992).
\end{references}
\end{document}